\documentclass{cpbtex}

\begin{document}

\title{Quantum anomalous Hall effect with tunable Chern numbers induced by a $d$-wave sublattice-staggered altermagnetism}


\author{Lizhou Liu(刘立周)$^{1}$ and \ Qing-Feng Sun(孙庆丰)$^{1,2}$  \thanks{Corresponding author. E-mail:sunqf@pku.edu.cn}\\
$^{1}$ International Center for Quantum Materials, School of Physics,\\
Peking University, Beijing 100871, China\\
$^{2}${Hefei National Laboratory, Hefei 230088, China}}   


\date{\today}
\maketitle

\begin{abstract}
We construct a minimal spinful tight-binding model on a square lattice, where a $d$-wave sublattice-staggered altermagnetism drives the quantum anomalous Hall effect.
Here the exchange field is staggered between the two sublattices, where it takes opposite signs on $A$ and $B$ described by the Pauli matrix $\tau_z$.
The resulting insulating phases host tunable Chern numbers $\mathcal{C}=\pm1$ and $\mathcal{C}=\pm2$, controlled by the staggered exchange strength and the sublattice-staggered potential.
We determine the complete phase diagram, identify valley-resolved band inversions at the $X$ and $Y$ points in the Brillouin zone, and demonstrate chiral edge states together with quantized two-terminal conductance plateaus.
Our work provides a simple route to realizing the quantum anomalous Hall effect in compensated magnets via a $d$-wave sublattice-staggered altermagnetism.
\end{abstract}



\section{Introduction}
The quantum anomalous Hall effect (QAHE), characterized by dissipationless chiral edge states and quantized Hall conductance in the absence of external magnetic fields, has been at the forefront of topological condensed matter research \cite{Haldane1988,Hasan2010,Qi2011, Chang2023, Liu2016, Weng2015, Bansil2016, Ren2016, Xiao2018, Nagaosa2010, Xu2011, addwan2025, Mei2024, Li2020, Wang2025, Wu2024, Ren2024}. 
Conventional realizations of QAHE rely predominantly on systems with intrinsic ferromagnetism or magnetic doping, where the breaking of time-reversal symmetry is essential for inducing nonzero Chern numbers \cite{Chang2013,Qiao2010,Qiao2011, Yu2010, Wang2014, Sun2019, Wang2013, Liu2008}. 
While ferromagnetic approaches have seen experimental success, including the first experimental observation of QAHE in Cr-doped (Bi,Sb)$_2$Te$_3$ thin films \cite{Chang2013}, they often suffer from external magnetic field sensitivity, low operational temperatures, and limited tunability of the topological phase \cite{He2018}.

In contrast, compensated magnetic systems offer improved robustness and additional internal degrees of freedom for engineering topological states \cite{Smejkal2018,Otrokov2019}. 
Here, compensated magnets refer to magnetic states with vanishing net magnetization per unit cell, including both conventional antiferromagnets and altermagnets.
For example, magnetic topological insulators with layered antiferromagnetism have been proposed and realized as promising platforms for QAHE and axion insulators \cite{Mong2010,Deng2020}. 
Furthermore, several theoretical proposals have demonstrated that QAHE can emerge in compensated antiferromagnets, such as MnBi$_2$Te$_4$ and monolayer CrO \cite{Liang2025,Guo2023}.
These works establish antiferromagnetic-based QAHE as viable in principle, yet they often rely on delicate external tuning or specific crystal structures.

Recently, the concept of altermagnetism has been unveiled as a new class of collinear magnetism that combines compensated spin configurations with strongly spin-polarized band structures \cite{Smejkal2022, Smejkal2022a, Reichlova2024, Smejkal2022_review, Wang2024a}. 
Therefore, altermagnetism belongs to the broader family of compensated magnets. The essential distinction from a conventional Néel antiferromagnet is that, despite zero net magnetization, the magnetic space group of an altermagnet allows symmetry protected momentum dependent spin splitting, whereas Néel antiferromagnets often exhibit antiunitary symmetries that enforce spin degenerate bands.
Unlike conventional ferromagnets, altermagnets exhibit no net magnetization and thus generate no stray magnetic fields; and unlike antiferromagnets, they naturally host momentum-dependent spin splitting enforced by crystalline symmetries, which is robust across the entire Brillouin zone~\cite{Smejkal2022,Smejkal2022a}.
First-principles calculations have predicted several candidate altermagnetic materials, such as RuO$_2$, MnF$_2$ \cite{Smejkal2020,Yuan2020,add1}, while experimental evidence has already been reported in MnTe and MnTe$_2$ \cite{Zhu2024,Krempasky2024}. 
Owing to these advantages, altermagnetism provides a platform for spin-dependent transport phenomena \cite{Yi2025,Sun2025} and for hosting unconventional topological phases \cite{Wan2025,Wan2025a}. Examples include Majorana zero modes \cite{Ghorashi2024,Zhu2023,Li2023}, higher-order topological states \cite{Li2024,Ezawa2024,Liu2025HOWeyl}, and unconventional superconducting responses \cite{Sun2023,Papaj2023,Cheng2024a,Cheng2024b}.
Recent studies have shown that altermagnets, despite hosting zero net magnetization, can nevertheless exhibit anomalous Hall effect and even QAHE \cite{Sato2024,Attias2024,Li2025KM}, though typically only in the presence of strong electronic correlations or intricate spin-orbit coupling configurations.
Meanwhile, compensated altermagnetic states have been proposed where the altermagnetic exchange changes sign between sublattices in real space, thereby restoring spin degeneracy and suppressing the characteristic altermagnetic spin splitting \cite{Meier2025AAM}.
In the sublattice-staggered altermagnet, the altermagnetic exchange is compensated at the unit-cell level, while its sublattice-staggered sign structure imprints a sublattice-opposed spin texture in momentum space.
This is distinct from conventional antiferromagnets, where real-space compensation is typically accompanied by momentum-independent spin degeneracy, and from ordinary altermagnets, which allow symmetry-constrained momentum-space spin splitting but usually lack such a sublattice-opposed texture.
Nevertheless, it remains an open question whether altermagnetism, additional real-space inversion between sublattices, can still generate QAHE with a nonzero Chern number.

In this paper, we address this question by constructing a minimal square-lattice model with two sublattices and a $d$-wave sublattice-staggered altermagnetic exchange field.
Here ''sublattice-staggered'' is used only to indicate that the exchange field is reversed between the two sublattices, yielding a compensated magnetic configuration.
The model, illustrated in Fig.~\ref{fig1}, contains hopping and spin-orbit coupling terms that preserve time-reversal symmetry $\mathcal{T}$. We then introduce a $d$-wave sublattice-staggered altermagnetic field that breaks $\mathcal{T}$ and drives QAHE while keeping selected crystalline symmetries specified below.
We show that this system realizes multiple QAHE phases with tunable Chern numbers $\mathcal{C} =\pm1$ and $\mathcal{C} =\pm2$, controlled by the interplay between altermagnetic-field strength and sublattice-staggered potential. 
By analyzing bulk topology, phase diagrams, edge spectra, and quantized two-terminal conductance, we establish a complete picture of QAHE driven by a $d$-wave sublattice-staggered altermagnetic field.
Our results demonstrate that a $d$-wave sublattice-staggered altermagnetism provides a distinct route to QAHE in compensated magnets, where the $d$-wave form factor controls valley-resolved band inversions at $X$ and $Y$.

\begin{center}
\scalebox{0.4}{\includegraphics{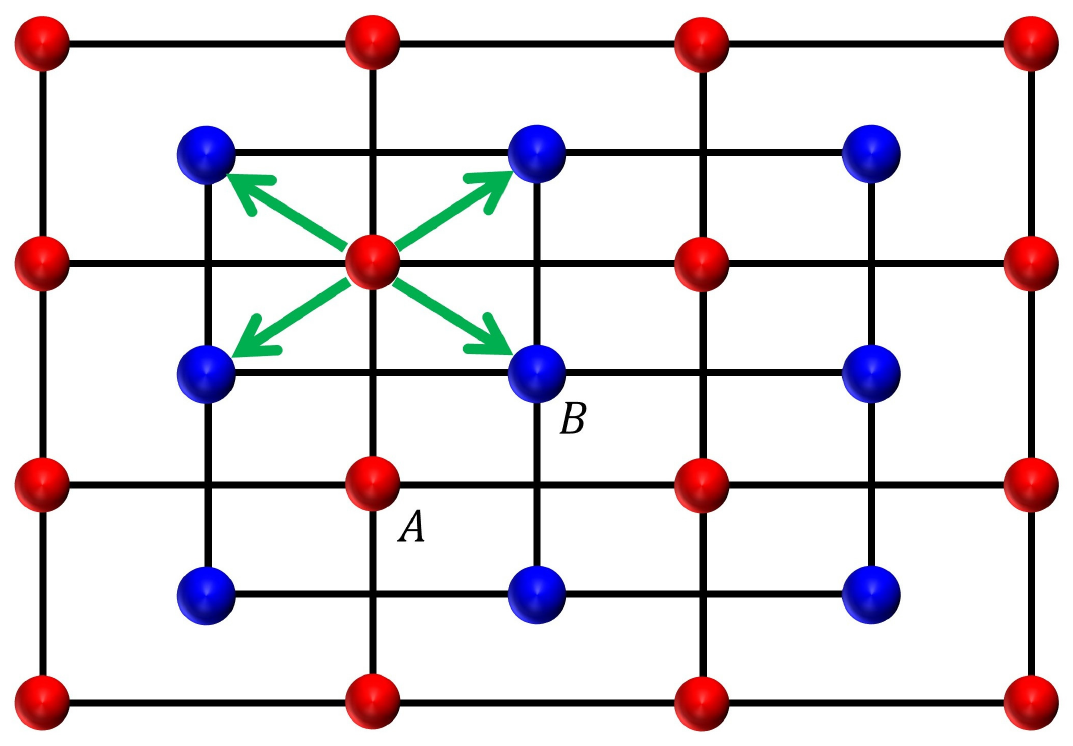}}\\[5pt]
\refstepcounter{figure}\label{fig1}
\parbox[c]{15cm}{\footnotesize{\bf Fig.~1.}   Schematic diagram of the square lattice model consisting of two sublattices, $A$ (red) and $B$ (blue). 
The green arrows indicate the diagonal hopping paths that include both nearest-neighbor hopping and spin-orbit coupling contributions. 
The $d$-wave staggered altermagnetism adds $+J(\cos k_x-\cos k_y)s_x$ on sublattice $A$ and $-J(\cos k_x-\cos k_y)s_x$ on sublattice $B$.
}
\end{center}

\section{Model and Hamiltonian}
We start from a spinful two-sublattice model defined on a square lattice. In momentum space, the Bloch Hamiltonian can be written as \cite{Guo2023}
\begin{equation}
H(\mathbf{k}) = 
\begin{bmatrix}
\Gamma^{A}_{\mathbf{k}} & \Gamma^{AB}_{\mathbf{k}} \\
\left( \Gamma^{AB}_{\mathbf{k}} \right)^\dagger & \Gamma^{B}_{\mathbf{k}}
\end{bmatrix},
\end{equation}
where the intra sublattice contributions are given by
\begin{align}
\Gamma^{A}_{\mathbf{k}} &= -2(t_x \cos k_x + t_y \cos k_y)\, s_0 + \mu\, s_0, \nonumber \\
\Gamma^{B}_{\mathbf{k}} &= 2(t_x \cos k_x + t_y \cos k_y)\, s_0 - \mu\, s_0.
\end{align}
The hoppings $(t_x,t_y)$ enter with opposite signs on sublattices $A$ and $B$, which favors an insulating spectrum once the inter-sublattice coupling is included.
The parameter $\mu$ is a sublattice-staggered potential, taking $+\mu$ on $A$ and $-\mu$ on $B$.
The inter-sublattice block $\Gamma^{AB}_{\mathbf{k}}$ describes the coupling between $A$ and $B$, including diagonal hopping $(t)$ and spin-orbit coupling $(i\lambda)$,
\begin{align}
\Gamma^{AB}_{\mathbf{k}} =\ & 2(t + i\lambda\, s_x)\cos\left( \frac{k_x + k_y}{2} \right) \nonumber \\
& + 2(t - i\lambda\, s_x)\cos\left( \frac{k_x - k_y}{2} \right).
\end{align} 
Here $s_i$ $(i=x,y,z)$ denote the Pauli matrices for spin, and $s_0$ is the $2\times2$ identity matrix. 

In the absence of spin-orbit coupling, the model has an enlarged symmetry that includes independent spin rotations combined with lattice operations \cite{Liu2022,Yang2024,Guo2021}.
Along the Brillouin zone boundaries ($k_x=\pi$ or $k_y=\pi$), $\Gamma^{AB}_{\mathbf{k}}$ vanishes, so bands originating from different sublattices cannot hybridize.
When spin-orbit coupling is switched on, the Hamiltonian remains time-reversal symmetric and does not imply magnetic order by itself.
Its role is to lock the spin quantization axis to lattice harmonics in the inter-sublattice coupling, thereby constraining the symmetry-allowed mass structures once a time-reversal-breaking exchange field is introduced.
Along the zone-boundary lines, the vanishing of $\Gamma^{AB}_{\mathbf{k}}$ yields symmetry-protected band touchings in the absence of exchange.
After adding the $d$-wave sublattice-staggered altermagnetism $H_{\mathrm{AM}}$ defined below, the system becomes magnetic, and the symmetry analysis is naturally formulated in terms of a magnetic point group~\cite{Guo2023}.
In this magnetic setting, the exchange generates momentum-dependent masses that gap out the relevant zone-boundary band touchings at $X$ and $Y$, producing a nontrivial Berry-curvature distribution and QAHE phases with tunable Chern numbers.

To fully gap out the system and realize QAHE, we introduce a $d$-wave sublattice-staggered altermagnetic exchange term,\begin{equation}
H_{\text{AM}}(\mathbf{k}) = J(\cos k_x - \cos k_y) \tau_z s_x .
\end{equation}
where $\tau_z$ acts on the sublattice degrees of freedom, such that the exchange contributes $+J(\cos k_x-\cos k_y)s_x$ on sublattice $A$ and $-J(\cos k_x-\cos k_y)s_x$ on sublattice $B$.
Here $s_x$ fixes the orientation of the staggered altermagnetic field along the $x$ direction.
Notably, when the staggered exchange field is aligned with the spin-orbit coupling spin axis, the combined effect produces masses of the same sign for the relevant zone-boundary valleys, opening a full bulk gap in the parameter regime of interest \cite{Guo2023}.
This term explicitly breaks time-reversal symmetry $\mathcal{T}$. 
We consider anisotropic hoppings $t_x$ and $t_y$.
The isotropic limit is recovered at $t_x=t_y$; otherwise $t_x\neq t_y$ describes hopping anisotropy.
The $d$-wave form factor $(\cos k_x-\cos k_y)$ vanishes on the diagonals $k_x=\pm k_y$. Nevertheless, the full bulk gap is determined by the complete Hamiltonian. In the parameter regime considered, the spin-orbit coupling together with $H_{\mathrm{AM}}$ gaps out the zone-boundary valleys at $X$ and $Y$, generating a nontrivial Berry curvature distribution and QAHE phases.
Compared with Ref. \cite{Guo2023}, which is based on a conventional antiferromagnetic exchange, our model instead employs a $d$-wave sublattice-staggered altermagnetic exchange field. 
This key modification produces a momentum-dependent exchange texture that is fundamentally different from a uniform staggering and directly targets the zone-boundary valleys at the $X$ and $Y$ points.
The $d$-wave momentum dependence further enables valley-resolved band inversions and hence tunable Chern numbers controlled by $J$, $\mu$, and the symmetry that relates the $X$ and $Y$ valleys.

\begin{center}
\scalebox{0.4}{\includegraphics{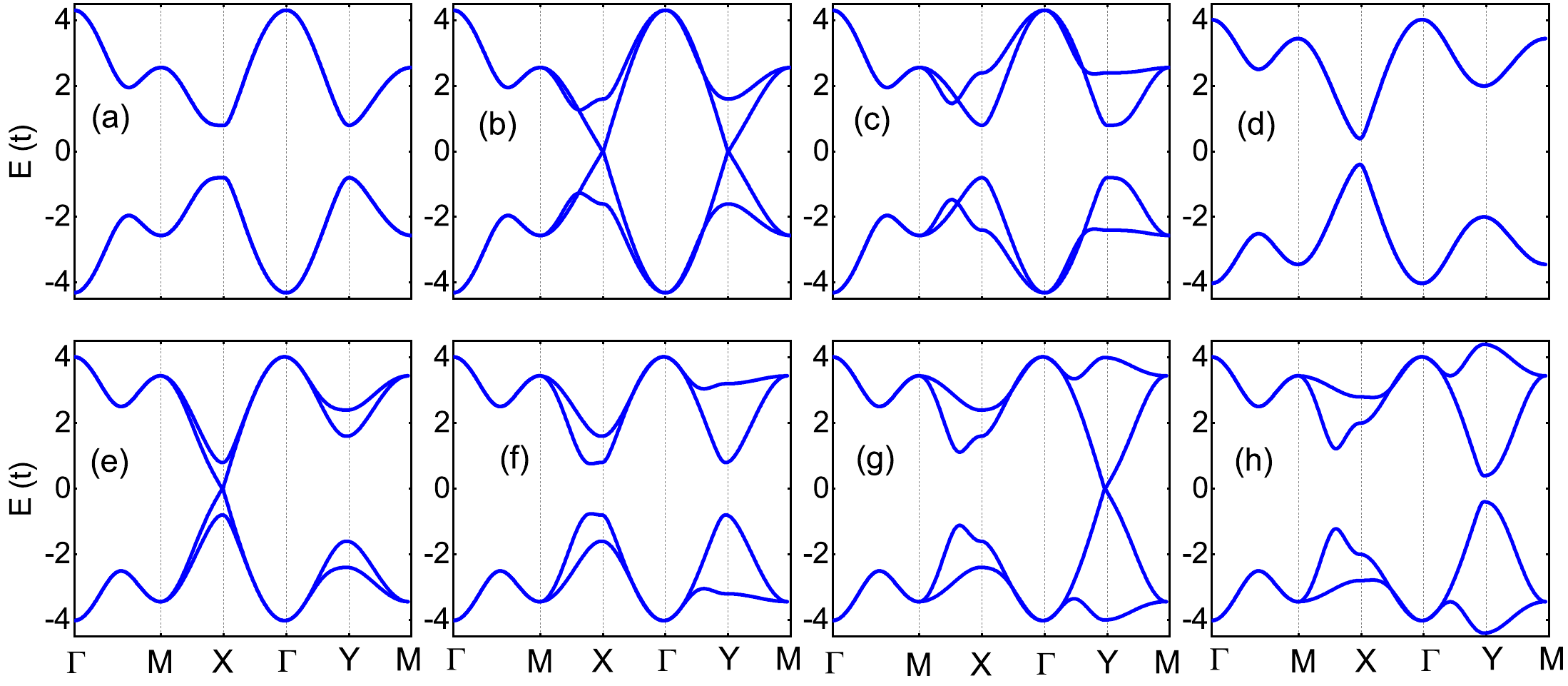}}\\[5pt]
\refstepcounter{figure}\label{fig2}
\parbox[c]{15cm}{\footnotesize{\bf Fig.~2.}  Energy band structures for different parameter settings. The high-symmetry path is chosen as $\Gamma$--M--X--$\Gamma$--Y--M. 
Panels (a)-(c) show the evolution at fixed $\mu = 0$ with increasing $J = 0$, 0.4, 0.8, where the bulk gap closes at $J = 0.4$, and the Chern number increases from $\mathcal{C} = 0$ to $\mathcal{C} = 2$. Panels (d)-(h) show results for $\mu = 1.2$ with increasing $J = 0, 0.2, 0.6, 1$, and $1.2$, where multiple topological transitions take place, with corresponding Chern numbers $\mathcal{C} = 0$, $1$, and $2$. 
Other parameters are fixed as $t = 1$, $\lambda = 0.5$, $t_x = 0.2$, $t_y = 0.6$.
}
\end{center}

\section{QAHE driven by a $d$-wave staggered altermagnetic exchange}
We now turn to the topological band evolution driven by the $d$-wave sublattice-staggered altermagnetism.
Figures~\ref{fig2}(a)--\ref{fig2}(c) display the band evolution in the absence of a sublattice-staggered potential ($\mu=0$). 
As $J$ increases from 0 to 0.8, the bulk gap closes simultaneously at the $X$ and $Y$ points when $J=0.4$, signaling a topological phase transition where two valley band inversions occur at the same critical point. Since each inversion contributes a unit change $\Delta\mathcal{C}=1$, the Chern number jumps from $\mathcal{C}=0$ to $\mathcal{C}=2$.
In contrast, for a finite staggered potential $\mu=1.2$, shown in Figs.~\ref{fig2}(d)--\ref{fig2}(h), the gap closes sequentially: first at $Y$ [Fig.~\ref{fig2}(e)] and later at $X$ [Fig.~\ref{fig2}(g)], giving rise to two successive transitions, each associated with a single valley inversion and a unit change $\Delta\mathcal{C}=1$, so that the Chern number evolves from $\mathcal{C}=0$ to $\mathcal{C}=1$ and finally to $\mathcal{C}=2$.
This comparison demonstrates that the staggered potential controls whether the band inversions at the $X$ and $Y$ valleys occur simultaneously or sequentially, and thus whether the Chern number changes by two in a single jump or by two unit steps, enabling QAHE phases with tunable Chern numbers.

The Chern number, serving as the topological invariant of the QAHE phases, is calculated using the Kubo formula~\cite{TKNN1982, Kohmoto1985}:
\begin{equation}
\mathcal{C} = \frac{1}{2\pi} \sum_n \int_{\mathrm{BZ}} d^2k\, \Omega_n(\mathbf{k}),
\end{equation}
where the Berry curvature \cite{Chang1996,Yao2004} of the $n$th band is
\begin{equation}
\Omega_n(\mathbf{k}) = - \sum_{n'\neq n} 
\frac{2\, \mathrm{Im}\!\left[\langle \psi_{n\mathbf{k}} | v_x | \psi_{n'\mathbf{k}} \rangle 
\langle \psi_{n'\mathbf{k}} | v_y | \psi_{n\mathbf{k}} \rangle \right]}
{(\omega_{n'} - \omega_n)^2},
\end{equation}
with $v_{x(y)}$ denoting the velocity operator and $\omega_n=E_n/\hbar$. 
Our calculations reproduce the Chern numbers identified in Fig.~\ref{fig2}, providing a consistent verification of the topological phase transitions.

The QAHE phases arise from the time-reversal breaking induced by $H_{\mathrm{AM}}$ together with SOC.
In the absence of $H_{\mathrm{AM}}$, the system respects $\mathcal{T}$ and the Berry curvature integrates to zero.
Turning on $H_{\mathrm{AM}}$ breaks $\mathcal{T}$ and generates finite Berry curvature, driving the system into QAHE phases with nonzero Chern numbers.
A finite sublattice-staggered potential $\mu$ shifts the relative energies of the $X$ and $Y$ valleys and can therefore turn a simultaneous band inversion into two sequential inversions, yielding intermediate $\mathcal{C}=\pm1$ phases when the symmetry relating $X$ and $Y$ is broken.
In particular, anisotropic hoppings $t_x\neq t_y$ remove the $k_x\leftrightarrow k_y$ interchange equivalence between the two directions, which makes the two valleys inequivalent and enables phases with odd Chern numbers.
In contrast, the isotropic limit $t_x=t_y$ restores the valley equivalence and restricts the phase diagram to even Chern numbers only.

\section{Phase diagram}
To gain a global view of the topological regimes, 
we construct the phase diagram in the $J$--$\mu$ plane by combining band-gap analysis with Chern number calculations, 
as shown in Fig.~\ref{fig3}. 
The color map represents the bulk gap magnitude. 
Figure~\ref{fig3}(a) corresponds to anisotropic hoppings $t_x\neq t_y$.
In this case, the system develops a rich phase structure with $\mathcal{C}=0,\pm1,\pm2$, 
originating from sequential gap closings at the $X$ and $Y$ valleys. 
In contrast, Fig.~\ref{fig3}(b) shows the isotropic hopping case $t_x=t_y$,
where the phase diagram simplifies dramatically and only $\mathcal{C}=0$ and $\mathcal{C}=\pm2$ phases appear.
This comparison indicates that the isotropic limit renders the $X$ and $Y$ valleys equivalent, enforcing simultaneous band inversions and hence even Chern numbers, whereas hopping anisotropy lifts this equivalence and allows valley-resolved inversions, leading to intermediate odd-Chern phases.

\begin{center}
\scalebox{0.4}{\includegraphics{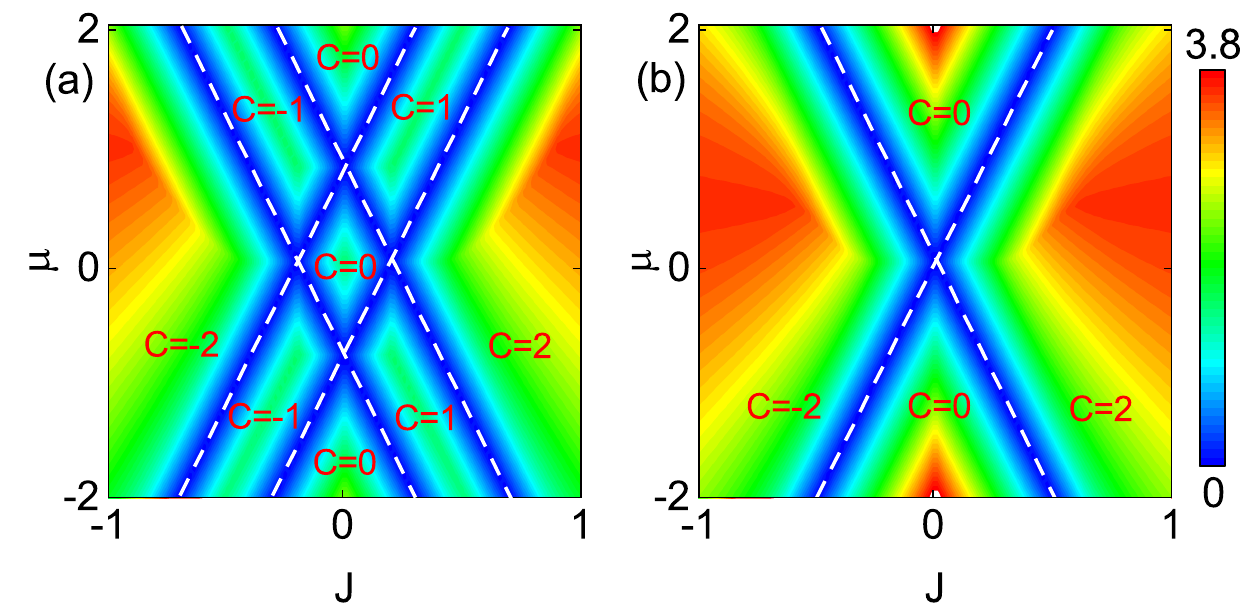}}\\[5pt]
\refstepcounter{figure}\label{fig3}
\parbox[c]{15cm}{\footnotesize{\bf Fig.~3.}  Phase diagrams in the $J$--$\mu$ plane. The color map shows the bulk gap magnitude, and the labeled regions indicate the corresponding Chern numbers. White dashed lines mark the phase boundaries.
Both panels share the common parameters: $t = 1$ and $\lambda = 0.5$. (a) $t_x = 0.2$, $t_y = 0.6$; (b) $t_x = 0.2$, $t_y = 0.2$.
}
\end{center}

To make the origin of the white dashed boundaries in Fig.~\ref{fig3} explicit, we derive the band-closing conditions from the low-energy expansions around the two zone-boundary valleys $X=(\pi,0)$ and $Y=(0,\pi)$.
Because the $d$-wave staggered exchange field
$H_{\mathrm{AM}}(\mathbf{k})=J(\cos k_x-\cos k_y)\tau_z s_x$
acts with opposite signs on the two sublattices, it modifies the diagonal blocks as
$\Gamma^{A}_{\mathbf{k}}\rightarrow \Gamma^{A}_{\mathbf{k}}+J(\cos k_x-\cos k_y)s_x$ and
$\Gamma^{B}_{\mathbf{k}}\rightarrow \Gamma^{B}_{\mathbf{k}}-J(\cos k_x-\cos k_y)s_x$.
Moreover, along the Brillouin-zone boundaries ($k_x=\pi$ or $k_y=\pi$), the inter-sublattice coupling $\Gamma^{AB}_{\mathbf{k}}$ vanishes identically, so the gap closing at $X$ and $Y$ is controlled by the diagonal sectors evaluated at the corresponding valley momenta.

We first expand around $X=(\pi,0)$ by setting $k_x=\pi+q_x$ and $k_y=q_y$ with $q_x,q_y\ll 1$.
Using $\cos k_x\simeq -(1-q_x^2/2)$ and $\cos k_y\simeq 1-q_y^2/2$, one has $\Gamma^{AB}_{\mathbf{k}}=0$ at $q=0$.
The Hamiltonian then reduces to two decoupled $2\times 2$ spin blocks on the two sublattices,
\begin{align}
\Gamma^{A}_{X}(0) &= 2(t_x-t_y)s_0+\mu s_0-2J s_x, \\
\Gamma^{B}_{X}(0) &= -2(t_x-t_y)s_0-\mu s_0+2J s_x .
\end{align}
Diagonalizing the spin structure yields $E=\pm\lvert \mu+2(t_x-t_y)\mp 2J\rvert$, so the gap closes at $X$ when
\begin{equation}
\lvert \mu+2(t_x-t_y)\rvert = 2J
\quad \Longrightarrow \quad
\mu = -2(t_x-t_y)\pm 2J .
\label{eq:boundary_X_main}
\end{equation}

A completely analogous expansion around $Y=(0,\pi)$ with $k_x=q_x$ and $k_y=\pi+q_y$ gives, at $q=0$,
\begin{align}
\Gamma^{A}_{Y}(0) &= -2(t_x-t_y)s_0+\mu s_0+2J s_x, \\
\Gamma^{B}_{Y}(0) &= 2(t_x-t_y)s_0-\mu s_0-2J s_x ,
\end{align}
leading to the band-closing condition
\begin{equation}
\lvert \mu-2(t_x-t_y)\rvert = 2J
\quad \Longrightarrow \quad
\mu = 2(t_x-t_y)\pm 2J .
\label{eq:boundary_Y_main}
\end{equation}
Equations~(\ref{eq:boundary_X_main}) and (\ref{eq:boundary_Y_main}) provide the analytical phase boundaries associated with gap closings at $X$ and $Y$, respectively, and they coincide with the white dashed lines in Fig.~\ref{fig3}.
This establishes a transparent valley interpretation of the phase diagram: Chern-number changes are triggered by gap closings at the zone-boundary valleys, while hopping anisotropy enters through the offset $2(t_x-t_y)$ that splits the $X$ and $Y$ closing lines and hence separates the two valley inversions.
As a result, the system can undergo two successive topological transitions with an intermediate odd-Chern regime.
In contrast, in the isotropic limit $t_x=t_y$ the offset vanishes and the two conditions collapse to $\mu=\pm 2J$, enforcing a simultaneous closing at $X$ and $Y$ and leaving only even-Chern phases.

\section{Chiral edge states and quantized transport}

To verify chiral edge states, we compute the ribbon spectrum with periodic boundary conditions along $x$ and open boundary conditions along $y$, so that $k_x$ remains a good quantum number.
Figures~\ref{fig4}(a) and \ref{fig4}(b) show the band structures for two representative parameter sets corresponding to Chern numbers $\mathcal{C} = 1$ and $\mathcal{C} = 2$, respectively. 
In both cases, gapless edge modes are clearly visible within the bulk energy gap.

\begin{center}
\scalebox{0.4}{\includegraphics{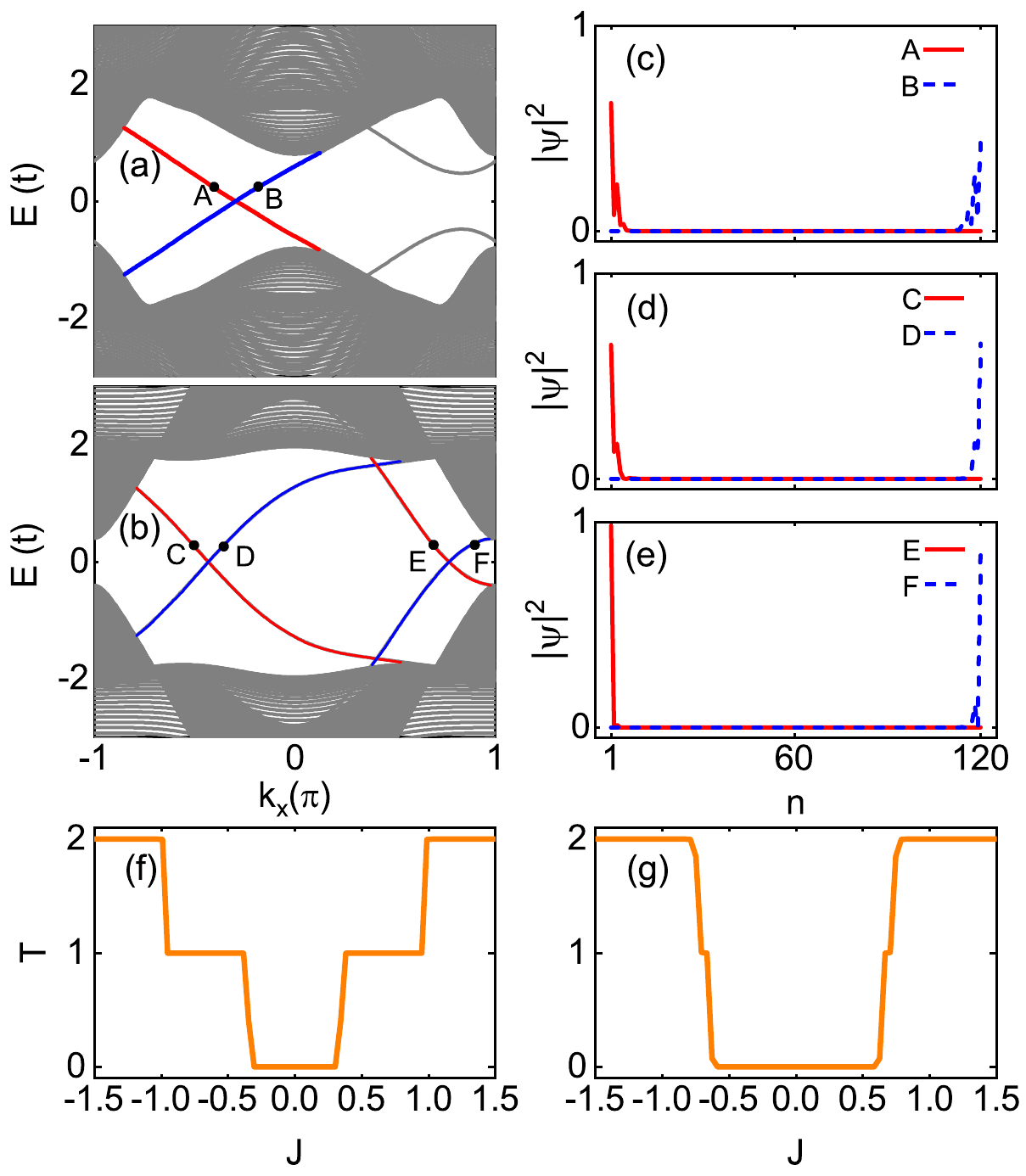}}\\[5pt]
\refstepcounter{figure}\label{fig4}
\parbox[c]{15cm}{\footnotesize{\bf Fig.~4.}  (a),(b) Ribbon band structures under open boundary conditions along the \(y\) direction for anisotropic hopping (\(t_x=0.2, t_y=0.6\)) at \(J=0.6\) and \(J=1.2\), corresponding to Chern numbers \(\mathcal{C}=1\) and \(\mathcal{C}=2\), respectively. 
(c)--(e) Real-space probability distributions of the edge states labeled $A--F$ in (a),(b), showing their localization at opposite edges. 
(f),(g) Two-terminal conductance as a function of altermagnetic strength \(J\) at fixed Fermi energy \(E=0.01\). 
Approximately quantized plateaus of \(0\), \(e^2/h\), and \(2e^2/h\) in (f) reflect successive transitions with \(\mathcal{C}=0,\pm1,\pm2\), while in (g) (isotropic hopping, \(t_x=t_y=0.2\)) the conductance jumps directly between \(0\) and \(2e^2/h\). 
Other parameters: \(t=1\), \(\lambda=0.5\), \(\mu=1.2\).
}
\end{center}

To examine their spatial localization, we select representative states labeled $A--F$ (black dots) and plot their real-space probability distributions in Figs.~\ref{fig4}(c), \ref{fig4}(d), and \ref{fig4}(f). 
For the $\mathcal{C} = 1$ phase [Fig.~\ref{fig4}(a)], the states $A$ and $B$ shown in Fig.~\ref{fig4}(c) localize at opposite edges, forming a single pair of chiral edge modes. 
In contrast, for the $\mathcal{C} = 2$ phase [Fig.~\ref{fig4}(b)], states $C–D$ and $E–F$ are both localized at the two edges of the ribbon. 
Importantly, states $C$ and $E$ propagate in the same direction and reside on the same edge, giving rise to two co-propagating chiral modes per edge~\cite{Qiao2010}. 
This direct correspondence between the number of edge branches and the bulk Chern number provides compelling evidence that the $d$-wave sublattice-staggered altermagnetism drives QAHE phases in our model.

To further confirm this bulk-boundary correspondence, we calculate the two-terminal conductance of a ribbon geometry using the Landauer-B$\rm \ddot{u}$ttiker formalism and the nonequilibrium Green's function approach \cite{Fisher1981,Metalidis2005,Sun2009,Lee1981}. 
In this setup, semi-infinite leads are attached to the left and right sides of the system, 
and the transmission coefficient is obtained as
$T(E)=\text{Tr}\!\left[\Gamma_L G^r \Gamma_R (G^r)^\dagger \right],$
where $\Gamma_{L(R)}(E)=i\left[\Sigma_{L(R)}^r- (\Sigma_{L(R)}^r)^\dagger \right]$ is the line-width function of the left (right) lead, $\Sigma^r_{L(R)}$ denotes the self-energy, and $G^{r}(E)=[E-H_C-\Sigma_L^{r}-\Sigma_R^{r}]^{-1}$ is the retarded Green's function of the central scattering region. The two-terminal conductance then follows as $G(E)=(e^2/h)T(E)$. 

At a fixed Fermi energy $E=0.01$, the conductance as a function of the altermagnetic strength $J$ is shown in Fig.~\ref{fig4}(f) and~\ref{fig4}(g). In the anisotropic hopping case [Fig.~\ref{fig4}(f)], corresponding to the phase diagram in Fig.~\ref{fig3}(a), distinct quantized plateaus of $0$, $e^2/h$, and $2e^2/h$ are observed, tracking successive topological transitions with $\mathcal{C}=0,\pm1,\pm2$. In contrast, the isotropic hopping case [Fig. \ref{fig4}(g)], associated with the phase diagram of Fig. \ref{fig3}(b), shows direct transitions between $\mathcal{C}=0$ and $\mathcal{C}=\pm2$, with conductance jumping between $0$ and $2e^2/h$ without intermediate plateaus. 
This highlights the role of the $X$--$Y$ valley equivalence in the isotropic limit: when $t_x=t_y$, only even Chern numbers appear, whereas hopping anisotropy $t_x\neq t_y$ allows intermediate odd-$\mathcal{C}$ plateaus.
These results establish quantized conductance as a robust signature of the valley-resolved topological transitions driven by the anisotropic $d$-wave altermagnetic order.

\section{Conclusion}
We have proposed a minimal square-lattice model to realize QAHE driven by a $d$-wave sublattice-staggered altermagnetism.
The interplay between the $d$-wave staggered altermagnetism strength and the sublattice-staggered potential gives rise to tunable topological phases with Chern numbers $\mathcal{C}=\pm1, \pm2$. 
Through bulk band structures, phase diagrams, edge spectra, and transport calculations, we have established the topological nature of the phases and their dependence on microscopic symmetry and coupling strength.
Our results establish the $d$-wave sublattice-staggered altermagnetism mechanism as a distinct and versatile route to engineer QAHE in compensated magnetic systems.

On the experimental side, altermagnetic order has been firmly established in several material platforms and thin-film settings, including MnTe and CrSb, where photoemission spectroscopy has directly resolved characteristic, symmetry-allowed band signatures of altermagnetism \cite{Krempasky2024,Reimers2024}.
For RuO$_2$, magnetic circular dichroism in photoemission has provided evidence for time-reversal-symmetry breaking without net magnetization \cite{Fedchenko2024}, and pronounced anomalous transport responses together with thin-film compatibility have been reported in related experiments \cite{Yang2025,Zhou2025,Noh2025,Leiviska2024}.
These advances identify realistic parent altermagnets with sizable spin-orbit coupling and device compatibility, which can serve as a materials basis to engineer the $d$-wave sublattice-staggered altermagnetism required in our minimal model.

A concrete device-level route is to integrate a thin-film altermagnet with engineered interfacial inversion breaking and sublattice asymmetry.
Electrostatic gating tunes the chemical potential into the bulk gap.
A controllable sublattice-staggered potential can be realized via an asymmetric interface, a polar substrate, or a ferroelectric layer that breaks the equivalence of the two sublattices.
In parallel, uniaxial strain provides an independent knob to tune the hopping anisotropy between $t_x$ and $t_y$, thereby generating the desired $d$-wave form factor and separating the $X$ and $Y$ valley inversions.
Experimentally, the proposed QAHE can be most directly identified through transport measurements under gate and strain control.
The resulting benchmarks include zero-field quantized Hall plateaus $R_{xy}=h/(\mathcal{C}e^2)$ accompanied by suppressed $R_{xx}$, and a characteristic plateau sequence that evolves from a single $\mathcal{C}=0\rightarrow2$ jump to two successive unit steps $\mathcal{C}=0\rightarrow1\rightarrow2$ upon tuning strain or gate voltage.
Moreover, since the altermagnetic electronic structure is tied to the N\'eel-vector orientation, electrical control or switching of the N\'eel vector provides a route to reversibly manipulate the quantized Hall response in a compensated setting \cite{Zhang2025,Chen2025}.

%
%

\section*{Data availability statement}
The data that supports the findings of this study are available from the author upon reasonable request.

\addcontentsline{toc}{chapter}{Acknowledgment}
\section*{Acknowledgments}
This work was financially supported by the National Key R and D Program of China (Grant No. 2024YFA1409002), the National Natural Science Foundation of China (Grants No. 12374034 and No. 12547169), and the Innovation Program for Quantum Science and Technology (Grant No. 2021ZD0302403). 
We also acknowledge the High-performance Computing Platform of Peking University for providing computational resources.

\addcontentsline{toc}{chapter}{References}
\begingroup
\def\bibname{References}
\def\refname{References}


\begin{thebibliography}{99}\footnotesize
\itemsep=-3pt plus.2pt minus.2pt

\bibitem{Haldane1988} Haldane F D M 1988 \textit{Phys. Rev. Lett.} \textbf{61} 2015

\bibitem{Hasan2010} Hasan M Z and Kane C L 2010 \textit{Rev. Mod. Phys.} \textbf{82} 3045
\bibitem{Qi2011} Qi X L and Zhang S C 2011 \textit{Rev. Mod. Phys.} \textbf{83} 1057
\bibitem{Chang2023} Chang C Z, Liu C X and MacDonald A H 2023 \textit{Rev. Mod. Phys.} \textbf{95} 011002

\bibitem{Liu2016} Liu C X, Zhang S C and Qi X L 2016 \textit{Annu. Rev. Condens. Matter Phys.} \textbf{7} 301

\bibitem{Weng2015} Weng H, Yu R, Hu X, Dai X and Fang Z 2015 \textit{Adv. Phys.} \textbf{64} 227

\bibitem{Bansil2016} Bansil A, Lin H and Das T 2016 \textit{Rev. Mod. Phys.} \textbf{88} 021004
\bibitem{Ren2016} Ren Y, Qiao Z and Niu Q 2016 \textit{Rep. Prog. Phys.} \textbf{79} 066501


\bibitem{Xiao2018} Xiao D, Jiang J, Shin J H, Wang W, Wang F, Zhao Y F, Liu C, Wu W, Chan M H W, Samarth N and Chang C Z 2018 \textit{Phys. Rev. Lett.} \textbf{120} 056801

\bibitem{Nagaosa2010} Nagaosa N, Sinova J, Onoda S, MacDonald A H and Ong N P 2010 \textit{Rev. Mod. Phys.} \textbf{82} 1539
\bibitem{Xu2011} Xu G, Weng H, Wang Z, Dai X and Fang Z 2011 \textit{Phys. Rev. Lett.} \textbf{107} 186806

\bibitem{addwan2025}
Wan Y H, Liu P Y, and Sun Q F
2025 \textit{Phys. Rev. Lett.} \textbf{135} 186302

\bibitem{Mei2024} Mei R, Zhao Y F, Wang C, Ren Y, Xiao D, Chang C Z and Liu C X 2024 \textit{Phys. Rev. Lett.} \textbf{132} 066604

\bibitem{Li2020} Li Z and Wang Z F 2020 \textit{Chin. Phys. B} \textbf{29} 107101


\bibitem{Wang2025} Wang W X, Liu Y W and He L 2025 \textit{Chin. Phys. B} \textbf{34} 047301


\bibitem{Wu2024} Wu X-C, Li S-Z, Si J-S, Huang B and Zhang W-B 2024 \textit{Chin. Phys. Lett.} \textbf{41} 057303

\bibitem{Ren2024} Ren X-L and Zhang C-W 2024 \textit{Chin. Phys. B} \textbf{33} 067102



\bibitem{Qiao2011} Qiao Z, Tse W K, Jiang H, Yao Y G and Niu Q 2011 \textit{Phys. Rev. Lett.} \textbf{107} 256801

\bibitem{Qiao2010} Qiao Z, Yang S A, Feng W, Tse W K, Ding J, Yao Y, Wang J and Niu Q 2010 \textit{Phys. Rev. B} \textbf{82} 161414(R)

\bibitem{Yu2010} Yu R, Zhang W, Zhang H J, Zhang S C, Dai X and Fang Z 2010 \textit{Science} \textbf{329} 61

\bibitem{Wang2014} Wang Q Z, Liu X, Zhang H J, Samarth N, Zhang S C and Liu C X 2014 \textit{Phys. Rev. Lett.} \textbf{113} 147201

\bibitem{Sun2019} Sun H, Xia B, Chen Z, Zhang Y, Liu P, Yao Q, Tang H, Zhao Y, Xu H and Liu Q 2019 \textit{Phys. Rev. Lett.} \textbf{123} 096401
\bibitem{Wang2013} Wang Z F, Liu Z and Liu F 2013 \textit{Phys. Rev. Lett.} \textbf{110} 196801


\bibitem{Liu2008} Liu C X, Qi X L, Dai X, Fang Z and Zhang S C 2008 \textit{Phys. Rev. Lett.} \textbf{101} 146802


\bibitem{Chang2013} Chang C Z, Zhang J, Feng X, Shen J, Zhang Z, Guo M, Li K, Ou Y, Wei P, Wang L L, Ji Z Q, Feng Y, Ji S, Chen X, Jia J, Dai X, Fang Z, Zhang S C, He K, Wang Y, Lu L, Ma X C and Xue Q K 2013 \textit{Science} \textbf{340} 167

\bibitem{He2018} He K, Wang Y and Xue Q K 2018 \textit{Annu. Rev. Condens. Matter Phys.} \textbf{9} 329

\bibitem{Smejkal2018} \v{S}mejkal L, Mokrousov Y, Yan B and MacDonald A H 2018 \textit{Nat. Phys.} \textbf{14} 242

\bibitem{Otrokov2019} Otrokov M M, Klimovskikh I I, Bentmann H, Estyunin D, Zeugner A, Aliev Z S, Ga\ss{}, Wolter A U B, Koroleva A V, Shikin A M, Blanco-Rey M, Hoffmann M, Rusinov I P, Vyazovskaya A Yu, Eremeev S V, Koroteev Yu M, Kuznetsov V M, Freyse F, S\'anchez-Barriga J, Amiraslanov I R, Babanly M B, Mamedov N T, Abdullayev N A, Zverev V N, Alfonsov A, Kataev V, B\"uchner B, Schwier E F, Kumar S, Kimura A, Petaccia L, Di Santo G, Vidal R C, Schatz S, Ki\ss{}ner K, \"Unzelmann M, Min C H, Moser S, Peixoto T R F, Reinert F, Ernst A, Echenique P M, Isaeva A and Chulkov E V 2019 \textit{Nature} \textbf{576} 416

\bibitem{Mong2010} Mong R S K, Essin A M and Moore J E 2010 \textit{Phys. Rev. B} \textbf{81} 245209
\bibitem{Deng2020} Deng Y, Yu Y, Shi M Z, Guo Z, Xu Z, Wang J, Chen X H and Zhang Y 2020 \textit{Science} \textbf{367} 895

\bibitem{Liang2025} Liang W, Li Z, An J, Ren Y, Qiao Z and Niu Q 2025 \textit{Phys. Rev. Lett.} \textbf{134} 116603

\bibitem{Guo2023} Guo P J, Liu Z X and Lu Z Y 2023 \textit{npj Comput. Mater.} \textbf{9} 70

\bibitem{Smejkal2022} \v{S}mejkal L, Sinova J and Jungwirth T 2022 \textit{Phys. Rev. X} \textbf{12} 031042

\bibitem{Smejkal2022a} \v{S}mejkal L, Sinova J and Jungwirth T 2022 \textit{Phys. Rev. X} \textbf{12} 040501

\bibitem{Reichlova2024} Reichlov\'a H, Sinova J, \v{S}mejkal L, Jungwirth T, Krempask\'y J and Springholz G 2024 \textit{Science} \textbf{384} 545

\bibitem{Smejkal2022_review} \v{S}mejkal L, Gonz\'alez-Hern\'andez R, Jungwirth T and Sinova J 2022 \textit{Nat. Rev. Mater.} \textbf{7} 482

\bibitem{Wang2024a} Wang Q, Wu D W, Guo G-H, Long M Q and Wang Y P 2024 \textit{Chin. Phys. B} \textbf{33} 097507


\bibitem{Smejkal2020} \v{S}mejkal L, Gonz\'alez-Hern\'andez R, Jungwirth T and Sinova J 2020 \textit{Sci. Adv.} \textbf{6} eaaz8809

\bibitem{Yuan2020} Yuan L D, Wang Z, Luo J W, Rashba E I and Zunger A 2020 \textit{Phys. Rev. B} \textbf{102} 014422
\bibitem{add1} Ga Y, Zhang F, Wang L, Jiang J, Chang K and Yang H 2025 \textit{Phys. Rev. B} \textbf{112} L020407




\bibitem{Zhu2024} Zhu Y P, Chen X, Liu X R, Liu Y and Liu P 2024 \textit{Nature} \textbf{626} 523

\bibitem{Krempasky2024} Krempask\'y J, \v{S}mejkal L, D'Souza S W, Hajlaoui M, Springholz G and Jungwirth T 2024 \textit{Nature} \textbf{626} 517

\bibitem{Yi2025} Yi X J, Mao Y, Lu X and Sun Q F 2025 \textit{Phys. Rev. B} \textbf{111} 035423

\bibitem{Sun2025} Sun Y F, Mao Y, Zhuang Y C and Sun Q F 2025 \textit{Phys. Rev. B} \textbf{112} 094411
\bibitem{Wan2025} Wan Y H and Sun Q F 2025 \textit{Phys. Rev. B} \textbf{111} 045407

\bibitem{Wan2025a} Wan Y H, Liu P Y and Sun Q F 2025 \textit{Phys. Rev. B} \textbf{112} 115412
\bibitem{Ghorashi2024} Ghorashi S A A, Hughes T L and Cano J 2024 \textit{Phys. Rev. Lett.} \textbf{133} 106601

\bibitem{Zhu2023} Zhu D, Zhuang Z Y, Wu Z and Yan Z 2023 \textit{Phys. Rev. B} \textbf{108} 184505
\bibitem{Li2023} Li Y X and Liu C C 2023 \textit{Phys. Rev. B} \textbf{108} 205410

\bibitem{Li2024} Li Y X, Liu Y and Liu C C 2024 \textit{Phys. Rev. B} \textbf{109} L201109

\bibitem{Ezawa2024} Ezawa M 2024 \textit{Phys. Rev. B} \textbf{109} 245306

\bibitem{Liu2025HOWeyl} Liu L, Sun Q F and Zhang Y T 2025 \textit{arXiv} arXiv:2502.13535

\bibitem{Sun2023} Sun C, Brataas A and Linder J 2023 \textit{Phys. Rev. B} \textbf{108} 054511

\bibitem{Papaj2023} Papaj M 2023 \textit{Phys. Rev. B} \textbf{108} L060508
\bibitem{Cheng2024a} Cheng Q, Mao Y and Sun Q F 2024 \textit{Phys. Rev. B} \textbf{110} 014518
\bibitem{Cheng2024b} Cheng Q and Sun Q F 2024 \textit{Phys. Rev. B} \textbf{109} 024517

\bibitem{Attias2024} Attias L, Levchenko A and Khodas M 2024 \textit{Phys. Rev. B} \textbf{110} 094425

\bibitem{Sato2024} Sato T, Haddad S, Fulga I C, Assaad F F and van den Brink J 2024 \textit{Phys. Rev. Lett.} \textbf{133} 086503

\bibitem{Li2025KM} Li Z, Li Z and Qiao Z 2025 \textit{Phys. Rev. B} \textbf{111} 155303

\bibitem{Meier2025AAM} Meier Q N, Carta A, Ederer C and Cano A 2025 \textit{arXiv} arXiv:2502.01515

\bibitem{Liu2022} Liu P, Li J, Han J, Wan X and Liu Q 2022 \textit{Phys. Rev. X} \textbf{12} 021016

\bibitem{Yang2024} Yang J, Liu Z X and Fang C 2024 \textit{Nat. Commun.} \textbf{15} 10203

\bibitem{Guo2021} Guo P J, Wei Y W, Liu K, Liu Z X and Lu Z Y 2021 \textit{Phys. Rev. Lett.} \textbf{127} 176401

\bibitem{TKNN1982} Thouless D J, Kohmoto M, Nightingale M P and den Nijs M 1982 \textit{Phys. Rev. Lett.} \textbf{49} 405

\bibitem{Kohmoto1985} Kohmoto M 1985 \textit{Ann. Phys. (N.Y.)} \textbf{160} 343

\bibitem{Chang1996} Chang M C and Niu Q 1996 \textit{Phys. Rev. B} \textbf{53} 7010

\bibitem{Yao2004} Yao Y, Kleinman L, MacDonald A H, Sinova J, Jungwirth T, Wang D S, Wang E and Niu Q 2004 \textit{Phys. Rev. Lett.} \textbf{92} 037204


\bibitem{Fisher1981} Fisher D S and Lee P A 1981 \textit{Phys. Rev. B} \textbf{23} 6851

\bibitem{Metalidis2005} Metalidis G and Bruno P 2005 \textit{Phys. Rev. B} \textbf{72} 235304

\bibitem{Sun2009} Sun Q F and Xie X C 2009 \textit{J. Phys.: Condens. Matter} \textbf{21} 344204

\bibitem{Lee1981} Lee D H and Joannopoulos J D 1981 \textit{Phys. Rev. B} \textbf{23} 4997


\bibitem{Krempasky2024} Krempask\'y J, \v{S}mejkal L, D'Souza S W, Hajlaoui M, Springholz G, Uhl\'i\v{r}ov\'a K, Alarab F, Constantinou P C, Strocov V N, Usanov D, Pudelko W R, Gonz\'alez-Hern\'andez R, Hellenes A B, Jansa Z, Reichlov\'a H, \v{S}ob\'{a}\v{n} Z, Gonzalez Betancourt R D, Wadley P, Sinova J, Kriegner D, Min\'ar J, Dil J H and Jungwirth T 2024 \textit{Nature} \textbf{626} 517

\bibitem{Reimers2024} Reimers S, Odenbreit L, \v{S}mejkal L, Strocov V N, Constantinou P, Hellenes A B, Ubiergo R J, Campos W H, Bharadwaj V K, Chakraborty A, Denneulin T, Shi W, Dunin-Borkowski R E, Das S, Kl\"aui M, Sinova J and Jourdan M 2024 \textit{Nat. Commun.} \textbf{15} 2116

\bibitem{Fedchenko2024} Fedchenko O, Min\'ar J, Akashdeep A, D'Souza S W, Vasilyev D, Tkach O, Odenbreit L, Nguyen Q, Kutnyakhov D, Wind N, Wenthaus L, Scholz M, Rossnagel K, Hoesch M, Aeschlimann M, Stadtm\"uller B, Kl\"aui M, Sch\"onhense G, Jungwirth T, Hellenes A B, Jakob G, \v{S}mejkal L, Sinova J and Elmers H J 2024 \textit{Sci. Adv.} \textbf{10} eadj4883



\bibitem{Yang2025} Yang G, Li Z, Yang S, Li J, Zheng H, Zhu W, Pan Z, Xu Y, Cao S, Zhao W, Jana A, Zhang J, Ye M, Song Y, Hu L-H, Yang L, Fujii J, Vobornik I, Shi M, Yuan H, Zhang Y, Xu Y and Liu Y 2025 \textit{Nat. Commun.} \textbf{16} 1442

\bibitem{Zhou2025} Zhou Z, Cheng X, Hu M, Chu R, Bai H, Han L, Liu J, Pan F and Song C 2025 \textit{Nature} \textbf{638} 645

\bibitem{Noh2025} Noh S, Kim G H, Lee J, Jung H, Seo U, So G, Lee J, Lee S, Park M, Yang S, Oh Y, Jin H, Sohn C and Yoo J W 2025 \textit{Phys. Rev. Lett.} \textbf{134} 246703




\bibitem{Leiviska2024} Leivisk\"a M, Rial J, Bad'ura A, Seeger R L, Kouta I, Beckert S, Kriegner D, Joumard I, Schmoranzerov\'a E, Sinova J, Gomonay O, Thomas A, Goennenwein S T B, Reichlov\'a H, \v{S}mejkal L, Michez L, Jungwirth T and Baltz V 2024 \textit{Phys. Rev. B} \textbf{109} 224430





\bibitem{Zhang2025} Zhang Y, Bai H, Dai J, Han L, Chen C, Liang S, Cao Y, Zhang Y, Wang Q, Zhu W, Pan F and Song C 2025 \textit{Nat. Commun.} \textbf{16} 5646

\bibitem{Chen2025} Chen Y, Liu X, Lu H Z and Xie X C 2025 \textit{Phys. Rev. Lett.} \textbf{135} 016701



\end{thebibliography}
\end{document}